\documentclass[letterpaper]{article} 
\usepackage{aaai21}  
\usepackage{times}  
\usepackage{helvet} 
\usepackage{courier}  
\usepackage[hyphens]{url}  
\usepackage{graphicx} 
\usepackage{natbib}  
\usepackage{caption} 
\usepackage{amsmath,amsfonts,amssymb}
\usepackage{prettyref}
\usepackage{mathrsfs}
\usepackage{graphicx}
\usepackage{wrapfig}
\usepackage{subfig}
\usepackage{sidecap}
\usepackage{MnSymbol}
\usepackage{multirow}
\usepackage{booktabs}
\usepackage{algorithm}
\usepackage{algpseudocode}
\usepackage[table]{xcolor}
\usepackage{cellspace}
\usepackage{mathtools}
\usepackage{comment}

\algnewcommand{\LineComment}[1]{\State \(\triangleright\) #1}
\algdef{SE}[DOWHILE]{Do}{doWhile}{\algorithmicdo}[1]{\algorithmicwhile\ #1}
\newcommand*{\colorboxed}{}
\def\colorboxed#1#{%
  \colorboxedAux{#1}%
}
\newcommand*{\colorboxedAux}[3]{%
  \begingroup
    \colorlet{cb@saved}{.}%
    \color#1{#2}%
    \boxed{%
      \color{cb@saved}%
      #3%
    }%
  \endgroup
}

\newrefformat{Fig}{Figure~\ref{#1}}
\newrefformat{fig}{Figure~\ref{#1}}
\newrefformat{par}{Section~\ref{#1}}
\newrefformat{appen}{Appendix~\ref{#1}}
\newrefformat{sec}{Section~\ref{#1}}
\newrefformat{sub}{Section~\ref{#1}}
\newrefformat{table}{Table~\ref{#1}}
\newrefformat{alg}{Algorithm~\ref{#1}}
\newrefformat{Alg}{Algorithm~\ref{#1}}
\newrefformat{Def}{Definition~\ref{#1}}
\newrefformat{Thm}{Theorem~\ref{#1}}
\newrefformat{Lem}{Lemma~\ref{#1}}
\newrefformat{step}{Step~\ref{#1}}
\newrefformat{ln}{Line~\ref{#1}}
\newrefformat{eq}{Equation~\ref{#1}}
\newrefformat{pb}{Problem~\ref{#1}}
\newrefformat{it}{Item~\ref{#1}}
\newrefformat{te}{Term~\ref{#1}}
\def\Eqref Eq:#1:{\eqref{eq:#1}}
\newrefformat{Eq}{Equation~\Eqref#1:}

\newcommand{\E}[1]{\mathbf{#1}}
\newcommand{\TE}[1]{\textbf{#1}}

\newcommand{\TWOR}[2]{\left(\setlength{\arraycolsep}{1pt}\begin{array}{cc}{#1}^T, & {#2}^T\end{array}\right)^T}
\newcommand{\THREE}[3]{\left(\setlength{\arraycolsep}{1pt}\begin{array}{ccc}{#1}, & {#2}, & {#3}\end{array}\right)}

\newcommand{\FOUR}[4]{\left(\setlength{\arraycolsep}{1pt}\begin{array}{cccc}{#1}, & {#2}, & {#3}, & {#4}\end{array}\right)}

\newcommand{\FOURR}[4]{\left(\setlength{\arraycolsep}{1pt}\begin{array}{cccc}{#1}^T, & {#2}^T, & {#3}^T, & {#4}^T\end{array}\right)^T}

\newcommand{\argmin}[1]{\underset{#1}{\E{argmin}}\;}
\newcommand{\argminP}[1]{\E{argmin}\;}

\newcommand{\argmaxP}[1]{\E{argmax}\;}
\newcommand{\ST}{\E{s.t.}\;}

\title{LCollision: Fast Generation of Collision-Free Human Poses\\
using Learned Non-Penetration Constraints}
\author {
        Qingyang Tan\textsuperscript{\rm 1},
        Zherong Pan\textsuperscript{\rm 2},
        Dinesh Manocha\textsuperscript{\rm 1} \\
}
\affiliations {
    \textsuperscript{\rm 1} Department of Computer Science, University of Maryland at College Park \\
    \textsuperscript{\rm 2} Department of Computer Science, University of Illinois at Urbana-Champaign \\
    qytan@umd.edu, zherong@illinois.edu, dmanocha@umd.edu\\
    \ \\
}

\begin{document}

\maketitle

\begin{abstract}
We present LCollision, a learning-based method that synthesizes collision-free 3D human poses. At the crux of our approach is a novel deep architecture that simultaneously decodes new human poses from the latent space and predicts colliding body parts. These two components of our architecture are used as the objective function and surrogate hard constraints in a constrained optimization for collision-free human pose generation. A novel aspect of our approach is the use of a bilevel autoencoder that decomposes whole-body collisions into groups of collisions between localized body parts. By solving the constrained optimizations, we show that a significant amount of collision artifacts can be resolved. Furthermore, in a large test set of $2.5\times 10^6$ randomized poses from SCAPE, our architecture achieves a collision-prediction accuracy of $94.1\%$ with $80\times$ speedup over exact collision detection algorithms. To the best of our knowledge, LCollision is the first approach that accelerates collision detection and resolves penetrations using a neural network.
\end{abstract}

\section{Introduction}
There has been considerable work on developing learning algorithms for 3D objects represented as point clouds \cite{qi2017pointnet}, meshes \cite{hanocka2019meshcnn}, volumetric grids \cite{Wang-2020-Completion}, and physical objects \cite{li2018learning}. Because these algorithms are used for different applications, a major challenge is accounting for user requirements and physics-based constraints. Considering these constraints can significantly improve the test-time robustness by preserving some known criteria of ``correct'' predictions. For example, we need to consider various forces and dynamics constraints for differentiable simulation~\cite{qiao2020scalable} and cloth embedding~\cite{tan2020cloth}, and a reliable robot motion planner should preserve a clearance distance from obstacles \cite{8460547}.

In this paper, we tackle the problem of collision-free human pose generation. Recently, 3D mesh representations have been used for learning-based human pose synthesis \cite{tretschk2019demea,bouritsas2019neural,ranjan2018generating,bagautdinov2018modeling,tan2018variational}. These methods learn a manifold of plausible human poses from a dataset, represented as the latent space of a deep autoencoder. Such autoencoders can be trained for applications including interactive rigging, human pose recognition from images and videos, and VR games. However, current learning-based methods do not account for any physics-based requirements such as (self-)collision-free constraint, thereby resulting in penetrations or other artifacts~\cite{tretschk2019demea,bouritsas2019neural,ranjan2018generating,bagautdinov2018modeling,tan2018variational}. By comparison, non-learning-based methods for character rigging~\cite{shi2007mesh} and physics-based simulation~\cite{Barbic:2010:SSC} can detect and explicitly handle the collisions using numerical methods. Our goal is to equip learned-based methods with similar collision-handling capabilities.

Although 3D data representations explicitly allow the modeling of collision-free constraints, satisfying these hard constraints in an end-to-end learning system is an open problem. Prior works have tried one of three ways to incorporate hard constraints in a learning system. First, classical second-order methods \cite{boggs1995sequential} for constrained optimization can enforce exact hard constraints on the parameters of the neural network. Second, variants of the stochastic projected gradient descend~\cite{MarquezNeila:262884,kervadec2019constrained} have been proposed to approximately satisfy the constraints on the neural network parameters. Finally, differentiable optimization layers \cite{8460547,agrawal2019differentiable} can modify the neural network output to satisfy such constraints. However, these methods are either limited to convex constraints, impractical for large networks, or do not provide sufficient accuracy in terms constraint satisfaction.

\TE{Main Results:} We present LCollision, a new learning algorithm to generate human poses that satisfy collision-free constraints. Our approach incorporates the non-penetration constraints by solving a general constrained optimization during the test time, where the feasible domain corresponding to these hard constraints is learned during the training time. The novel components of our approach include:
\begin{itemize}
\item \TE{Constrained Optimization Using Neural Network Function Approximation:} Instead of using exact collision-response, learning the feasible domain using a neural network provides approximate sub-gradients via back-propagation, which is much faster than exact collision-checking algorithms.
\item \TE{Collision Decomposition:} A collision only affects local regions of the human body, and we design our collision predictor to respect these local effects. Each point on the human body is softly assigned to a set of local body parts, and the collision loss is decomposed to these local domains, accordingly.
\item \TE{Hybrid Ranking, Potential Energy, and Entropy Loss:} Although exact hard constraints correspond to a binary loss (violation or non-violation), this loss should be differentiable so that constrained optimizations can be guided by gradient information. We propose a penetration-depth-based formulation~\cite{zhang2007generalized} as a collision metric to offer gradient direction, combined with the ranking loss to maintain the relative penetration depth between a pair of samples.
\end{itemize}
We have evaluated our method on 
the SCAPE dataset \cite{anguelov2005scape}, the MIT-Swing dataset \cite{vlasic2008articulated}, and the MIT Jumping dataset \cite{vlasic2008articulated}.
Combining these techniques, we achieve an accuracy of $94.1\%$, a false positive rate of $6.1\%$, and a false negative rate of $5.7\%$ when predicting collisions for $2.5\times 10^6$ randomized human poses sampled from these datasets. After learning the feasible domain, solving a constrained optimization for a collision-free human pose with $2161$ vertices takes $2.095$ iterations and $0.25$s on average. Moreover, our learned collision detector is $80\times$ faster than prior exact collision detection methods running on a CPU~\cite{pan2012fcl}.

\section{Related Work}
We review related works on human pose estimation and synthesis, collision detection and response, and deep network training with hard constraints.
\begin{figure*}[t]
\centering
\scalebox{0.95}{\includegraphics[width=0.98\linewidth]{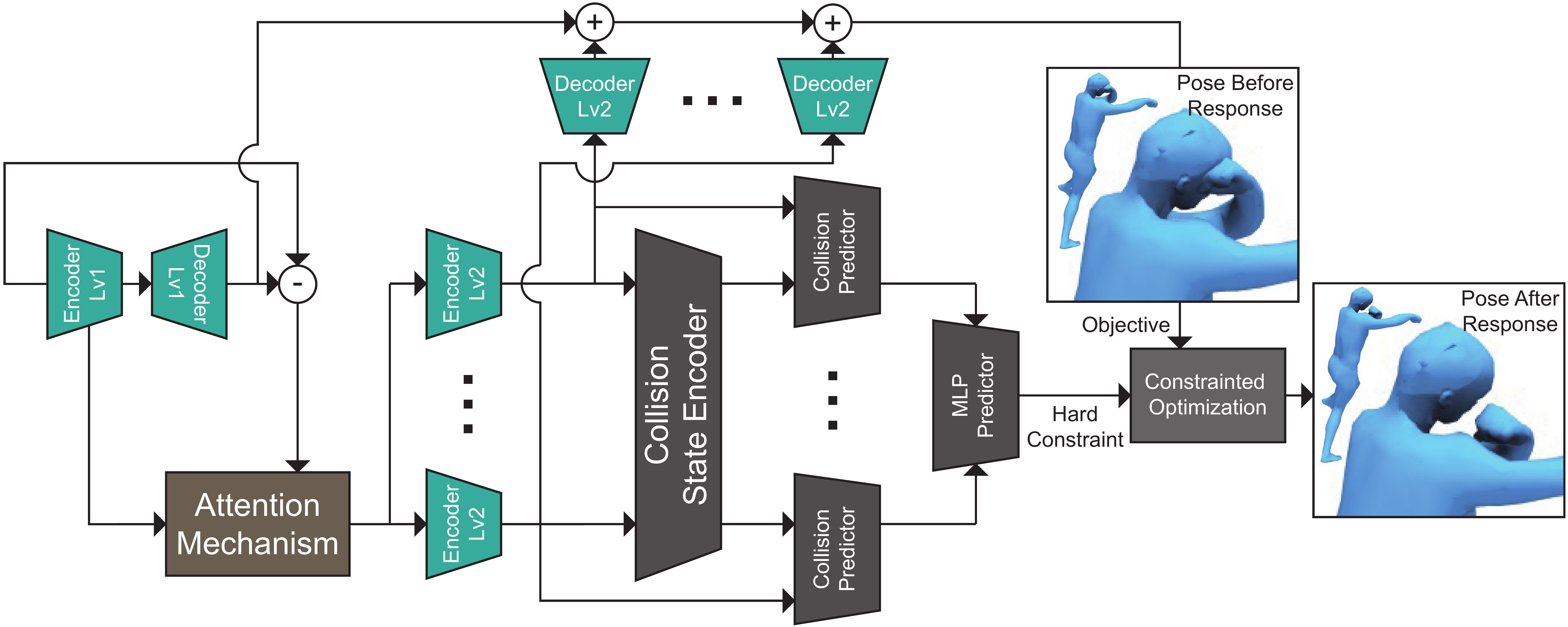}}
\put(-455,70){$\E{C}$}
\put(-383,170){$\bar{\mathcal{X}}_0$}
\put(-395,60){\rotatebox{90}{$\mathcal{X}_0-\bar{\mathcal{X}}_0$}}
\put(-303,92){$\mathcal{Z}_1$}
\put(-303,20){$\mathcal{Z}_{|\mathcal{Z}_0|}$}
\put(-248,92){$\mathcal{S}_0$}
\put(-248,20){$\mathcal{S}_0$}
\put(-203,110){$\mathcal{S}_1$}
\put(-203,20){$\mathcal{S}_{|\mathcal{Z}_0|}$}
\vspace{-5px}
\caption{\label{fig:pipeline} Our network architecture combines the domain-decomposed human pose embedding framework (green) and a novel collision state estimator (gray). Given an input pose, we use a weight-shared, level-1 autoencoder to learn a global shape embedding. The error on each domain is further reduced using a set of level-2 autoencoders. Both the level-1 and level-2 autoencoders' latent codes are used to predict a global collision state. Finally, the latent code of each level-2 autoencoder is compared against the global collision state to infer a localized penetration depth. These inferred penetrations are used in hard constraints of a constrained optimization framework for collision handling.}
\vspace{-10px}
\end{figure*}
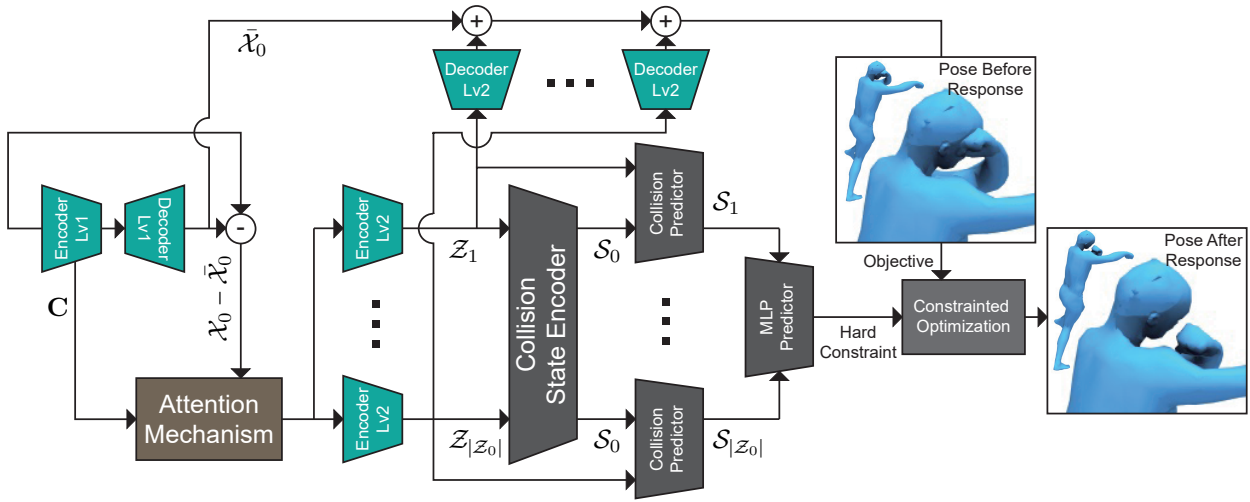

\TE{Human Pose Estimation \& Synthesis:} There is considerable work on human pose estimation and synthesis. Earlier methods~\cite{pedesrain2005} represent a pedestrian as a bounding box. An improved algorithm was proposed in \cite{agarwal2005recovering}, and this algorithm predicts the 55-D joint angles for a skeletal human pose. More accurate prediction results have been proposed in \cite{rogez2008randomized} using random forests and in \cite{toshev2014deeppose} using convolutional neural networks. Our approach is based on recent learning methods~\cite{tan2018variational,tretschk2019demea} that use 3D meshes to generate detailed human poses. Mesh-based representations
are inherently difficult to learn due to the intrinsic high-dimensionality, and the algorithm can produce sub-optimal results with various artifacts such as self-penetrations, noisy mesh surfaces, and flipped meshes. In view of these problems, \cite{villegas2018neural} only computes skeletal poses using learning and then uses skinning to recover the mesh-based representation. However, this approach requires additional skeleton-mesh correspondence information, which is typically unavailable in many datasets, including SCAPE \cite{anguelov2005scape}. 

\TE{Collision Detection \& Response:} An important criterion of ``correct'' human body shapes is that they are (self-) collision-free, i.e. elements of the mesh do not penetrate each other. Collision detection and response computations have been well-studied, with many practical algorithms proposed for large-scale 3D meshes \cite{pan2012fcl,kim2018chapter} that can be used to resolve penetrations. Collisions can be handled in a discrete or continuous manner. Discrete collision handling \cite{kim2018chapter} assumes that meshes can occasionally reach an invalid status with penetrations and therefore checks for collisions at fixed time intervals. In contrast, continuous collision detection algorithms estimate the time instance corresponding to the first contact and thereby maintain non-penetration configurations. These continuous collision detection (CCD) methods~\cite{tang2009iccd,bridson2002robust,tang2010fast} make some assumptions about the interpolating motion between two time instances and use analytic methods to predict the time of the collision. Many of these methods can be accelerated using GPU parallelism~\cite{govindaraju2005quick}. In theory, we can use different collision handling methods to avoid penetrations in a 3D mesh of a human pose. However, there are two practical challenges. First, collision response is a physical behavior tightly coupled with a physics-based model of the human body. However, modeling the physical deformations of a human body can be computationally expensive. The running time for simulating one timestep of a human body can be more than $20$ seconds~\cite{smith2018stable}, which is intractable for interactive applications. Second, collision handling algorithms require a volumetric mesh, while many applications of human pose synthesis rely on surface meshes. Techniques have also been proposed to estimate the extent of penetrations between complex 3D geometric models, e.g., penetration depth~\cite{zhang2007generalized,kim2002fast,6280111}. However, these formulations can be non-smooth and expensive to compute.

\TE{Training Deep Networks with Hard Constraints:} An additional layer of challenge is to incorporate collision handling into a deep learning framework. In particular, state-of-the-art deep learning methods are unable to handle such hard constraints. Constraints on neural network parameters \cite{ravi2019explicitly} are used for regularizing the network training and can be approximately enforced using variants of the projected gradient descend algorithm. On the other hand, constraints on neural network output are used to model application-specific requirements such as collision-free constraints. Prior works \cite{8460547,agrawal2019differentiable,MarquezNeila:262884,nandwani2019primal} use a similar approach to enforce hard constraints: converting the constrained optimization into an unconstrained min-max optimization, which can be solved approximately by updating the primal and dual variables. A special case arises when the hard constraints are convex; then the constrained optimization can be solved efficiently with exact constraint enforcement~\cite{8460547,agrawal2019differentiable}. However, the collision-free constraints in our applications are neither convex nor smooth.

\section{Human Pose \& Collision-Free Constraints}
Recent methods~\cite{tretschk2019demea,bouritsas2019neural,ranjan2018generating,bagautdinov2018modeling,tan2018variational} have used neural networks to generate new poses from a small set of examples via shape embedding. In this section, we give an overview of the process of computing the embedding space for human pose generation and highlight the collision-free constraints that LCollision tries to satisfy.

\subsection{Human Pose Embedding}
Our method uses the algorithm in \cite{tan2018mesh, yang2020multiscale}, which has the ability to extract local deformation components (more details given in the appendix). We represent human models as triangle meshes -- a special graph $\mathcal{G}=<\mathcal{V},\mathcal{E}>$, with $\mathcal{V}$ being a set of vertices and $\mathcal{E}$ being a set of edges. In our datasets, all the models share the same topology, i.e. $\mathcal{E}$ is the same over all the meshes, while $\mathcal{V}$ differs. We transform $\mathcal{V}$ to the as-consistent-as-possible (ACAP) feature space \cite{gao2019sparse}, denoted as $\mathcal{X}\in\mathbb{R}^{9\times|\mathcal{V}|}$, to  handle large deformations. We use a bilevel autoencoder to embed $\mathcal{X}$ in a latent space. Both levels of the autoencoder involve one graph convolutional layer and one fully connected layer. The fully-connected layer maps the feature to a $K$-dimensional latent code, with weights denoted as $\E{C}\in\mathbb{R}^{K\times9\times|\mathcal{V}|}$. A sparsity loss is used to ensure that each dimension of $\E{C}$ only accounts for a group of local points.

\begin{figure}[t]
\vspace{-5px}
\centering
\scalebox{0.8}{\includegraphics[width=0.95\linewidth]{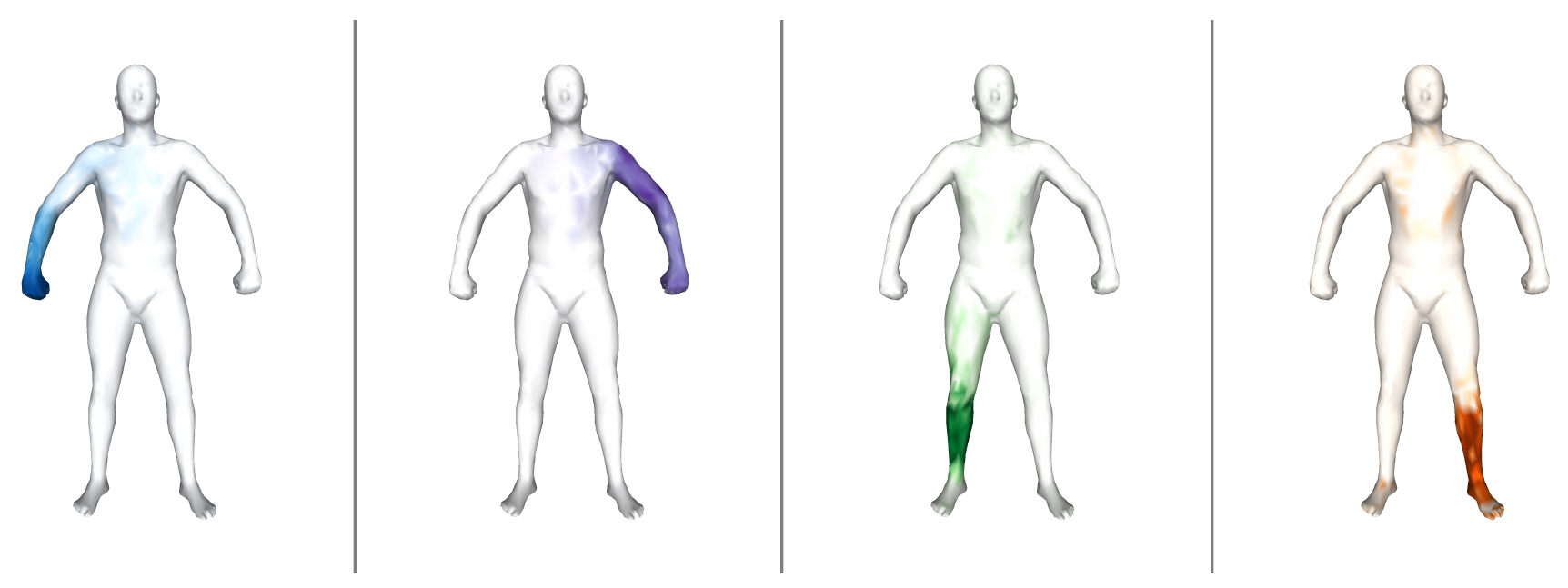}}
\vspace{-8px}
\caption{\label{fig:decompose} We show decomposed domains on the SCAPE dataset using the learned attention mask $\mathcal{M}^{ki}$, highlighted in different colors. The darkness of a given color represents the weight of the soft assignment. These weights are used for localized collision computations.}
\vspace{-10px}
\end{figure}

\TE{Domain Decomposition via Attention:}\label{sec:attention}
We use a bilevel architecture because we want the level-2 autoencoder to learn a decomposed domain of the original mesh, i.e. each level-2 autoencoder only reduces the level-1 residual on a subset of $\mathcal{V}$. The learned domain decomposition not only enhances the reusability and explainability of the neural network but is also used to model the local collisions between body sub-parts, as explained in \prettyref{sec:locallity}.

Each autoencoder maps some input feature $\mathcal{X}$ to a latent code $\mathcal{Z}$ and then reconstructs $\mathcal{Z}$ to feature $\bar{\mathcal{X}}$. We use subscripts to denote the index of an autoencoder, e.g., $\mathcal{X}_0$, $\mathcal{Z}_0$, and $\bar{\mathcal{X}}_0$ are the input, latent code, and output of the level-1 autoencoder, respectively. We assume that each entry of level-1 latent code corresponds to a sub-domain of the human body on which the residual is further reduced using one level-2 autoencoder, so there are altogether $|\mathcal{Z}_0|+1$ autoencoders. The $k$th level-2 autoencoder is responsible for representing a subset of residual $\mathcal{X}_0-\bar{\mathcal{X}}_0$. To determine the subset, an attention mask is computed as: $\mathcal{M}^{ki}={\sum_{j=1}^9{\E{C}^{kji}}^2}/{\sum_{k=1}^{|\mathcal{Z}_0|}\sum_{j=1}^9{\E{C}^{kji}}^2}$. In addition, the input to the $k$th level-2 autoencoder is $\mathcal{X}_k^i=\mathcal{M}^{ki}(\mathcal{X}_0^i-\bar{\mathcal{X}}_0^i)$. The soft assignment induced by the attention mask conducts the domain decomposition in our network. We illustrate some human body parts decomposed using $\mathcal{M}^{ki}$ in \prettyref{fig:decompose}. 

\subsection{Collision-Free Constraints}
A pivotal requirement of plausible human poses is that they are collision-free, i.e. triangles on the surface mesh do not penetrate each other. However, this constraint is ignored by previous neural-network-based human pose generation methods. We define a self-collision as an intersection between two topologically disjointed triangles, i.e. two triangles that do not share any edges. We use the following condition to indicate a collision: $\mathbf{t}_p \cap \mathbf{t}_q \neq \emptyset$, where $\mathbf{t}_p$ and $\mathbf{t}_q$ are two triangles.
Penetration depth (PD) is a notion that measures the extent of collision constraint violations between two objects. We define the local PD for triangle pair $(\mathbf{t}_p, \mathbf{t}_q)$ as:
\begin{align*}
\text{PD}_{p,q} = \min \{\|\mathbf{d}\|_2: (\mathbf{t}_p + \mathbf{d}) \cap \mathbf{t}_q = \emptyset \},
\end{align*}
where $\text{PD}_{p,q}$ is the minimum distance to move $\mathbf{t}_p$ such that $\mathbf{t}_p$ and $\mathbf{t}_q$ have no overlap. The collision-free constraint can be reformulated as the constraint that $\text{PD}_{p,q}=0$ for any $(p,q)$ pairs. Conceptually, collision constraints can be satisfied by solving the following constrained optimization:
\begin{equation*}
\begin{aligned}
\min&\quad goal \\ 
\ST& \text{PD}_{p,q} = 0,\quad (p,q)\;\;\text{disjoint},
\end{aligned}
\end{equation*}
where $goal$ is the objective (e.g., as close as possible to a user-desired pose). Prior works solve the constrained optimization by computing $\text{PD}_{p,q}$ for all $(p,q)$ pairs and treating each colliding $(\mathbf{t}_p, \mathbf{t}_q)$ as a standalone constraint, leading to large problem sizes and high computational costs.  Instead, we use a neural network to speed up the computation.

\subsection{\label{sec:locallity}Locality of Self-collisions}
Our method is inspired by the subspace self-collision culling algorithm (SSCC) \cite{Barbic:2010:SSC} and the learning-based collision simplification algorithm \cite{10.1145/2601097.2601181}. In SSCC, the authors observe that collisions usually occur between pairs of triangles that are originally close to one another on the template mesh. Pairs of distant triangles penetrate only when the mesh has undergone sufficient deformation. The observation made by SSCC suggests the use of mesh decompositions as described in \ref{sec:attention}. 

It is worth noting that both works~\cite{Barbic:2010:SSC,10.1145/2601097.2601181} use learned linear subspaces to accelerate collision detection and culling. However, the expressivity of linear subspaces is rather limited, so SSCC can only model deformations that are near the neutral pose and cannot represent larger deformations. Further, it is assumed in \cite{10.1145/2601097.2601181} that a domain decomposition is provided by users. Our work unifies and extends these ideas into a collision prediction algorithm that works for large deformations and does not require any additional data from users.

\begin{figure*}[t]
\centering
\scalebox{0.75}{\includegraphics[width=0.93\textwidth]{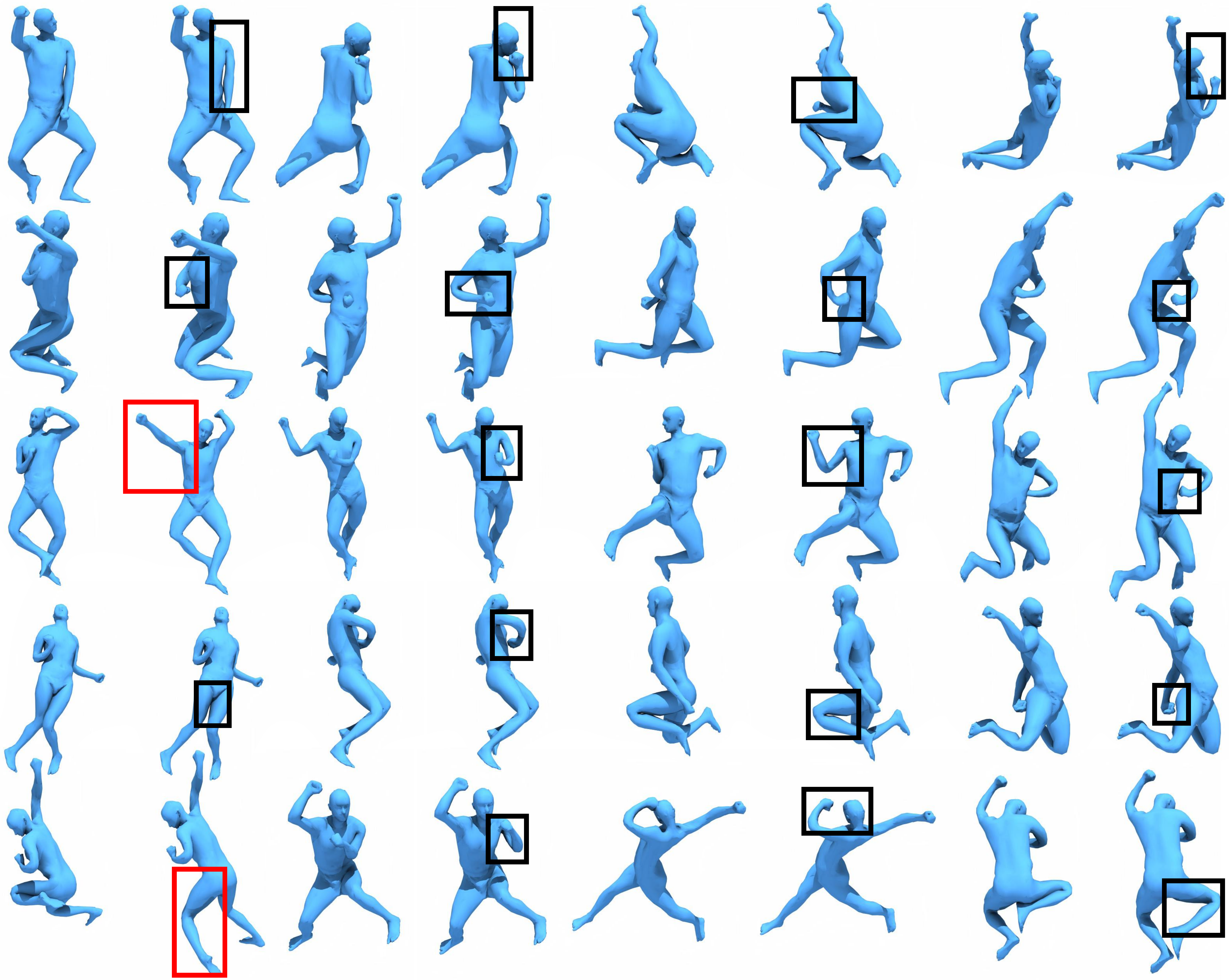}}
\vspace{-5px}
\caption{\small{\label{fig:all} We illustrate 20 representative results of collision responses, where the poses on the left are the original poses directly generated using \cite{yang2020multiscale}, and the poses on the right are the ones after collision responses. We highlight the adjusted body parts using black boxes. In all the examples, our method can successfully avoid penetrations. However, in two cases (red boxes), our adjusted poses drift severely from the original poses.}}
\vspace{-10px}
\end{figure*}

\section{LCollision: Overall Learning Algorithm}
Our overall learning architecture is illustrated in \prettyref{fig:pipeline}. Our method augments a normal mesh embedding autoencoder with an additional component to classify the collision status. Given a latent code $\mathcal{Z}_{all}$ defined as:
\begin{align*}
\mathcal{Z}_{all}=\FOURR{\mathcal{Z}_0}{\mathcal{Z}_1}{\cdots}{\mathcal{Z}_{|\mathcal{Z}_0|}},
\end{align*}
we output a collision probability $\text{MLP}_{classifier}$. We assume that the $0.5$ sub-level set of $\text{MLP}_{classifier}$ corresponds to collision-free meshes so that many constraints of the form $\text{PD}_{i,j}=0$ can be replaced by a single constraint $\text{MLP}_{classifier}<0.5$, which reduces the computational cost. 

\subsection{Collision Detection Architecture}
In this subsection, we explain our network architecture to cope with the locality of self-collisions illustrated in gray blocks of \prettyref{fig:pipeline}, including the collision state encoder and the collision predictor.

\TE{Naive Subdivision:} Our level-2 autoencoders inherently decompose the mesh into $|\mathcal{Z}_0|$ sub-domains. Therefore, if collisions occur within the $k$th sub-domain, then collisions should be inferred from $\mathcal{Z}_k$ alone, and we use a collision predictor (CP) in the form of a multilayer perceptron (MLP) to map $\mathcal{Z}_k$ to some collision indicator. If a pair of triangles belongs to two sub-domains, e.g., $\mathcal{Z}_k$ and $\mathcal{Z}_{k'}$, then a possible solution is to use another MLP that takes both $\TWOR{\mathcal{Z}_k}{\mathcal{Z}_{k'}}$. However, this approach requires $\mathcal{O}(|\mathcal{Z}_0|^2)$ CPs with an excessively large number of weights, and the latent codes of level-2 autoencoders only represent the relative residual $\mathcal{X}_0-\bar{\mathcal{X}}_0$, while the absolute information $\mathcal{X}_0$ is lost. 
\TE{Our Method:} To avoid issues with naive subdivision, we propose using a collision state encoder (CSE) that encodes both relative and absolute information over all mesh sub-domains. CSE is an MLP that takes $\mathcal{Z}_{all}$ and brings $\mathcal{Z}_{all}$ through three latent layers with $\THREE{512}{256}{256}$ neurons and ReLU activation. Finally, CSE outputs a latent code referred to as the global collision state, or $\mathcal{S}_0=\text{CSE}(\mathcal{Z}_{all})$ for short. $\mathcal{S}_0$ and $\mathcal{Z}_k$ are then fed into a CP to obtain the collision indicator related to the $k$th sub-domain, i.e. collisions between pairs of triangles where at least one of the triangles belongs to the $k$th sub-domain. There are altogether $|\mathcal{Z}_0|$ CPs, where the $k$th CP maps $\TWOR{\mathcal{S}_0}{\mathcal{Z}_k}$ through four latent layers with $\FOUR{512}{256}{256}{128}$ neurons and ReLU activation. Finally, CP outputs a scalar collision indicator $\mathcal{S}_k$, i.e. $\mathcal{S}_k=\text{CP}(\mathcal{S}_0,\mathcal{Z}_k)$.

\subsection{Collision Predictor Based on Penetration}\label{sec:collision_predict}
We need the collision indicators $\mathcal{S}_k$ and groundtruth labels $\mathcal{S}_i$ to be compatible with numerical optimization. Since we use gradient-based numerical optimization, we need to provide valid gradient information. To this end, $\mathcal{S}_k$ should not only be a collision indicator but also a collision violation metric. In other words, if $\mathcal{S}_k'>\mathcal{S}_k\geq0$, then we must have $\mathcal{S}_k'$ correspond to a mesh with more collisions than $\mathcal{S}_k$, for which we use the notion of penetration depth. Given a mesh $\mathcal{G}$, we use the FCL library \cite{pan2012fcl} to compute the squared penetration depth $\text{PD}_{p,q}^2$ of each colliding triangle pair. This colliding pair correlates 6 vertices in $\mathcal{V}$, and we add $\text{PD}_{p,q}^2/6$ to each vertex as the vertex-wise collision violation. After processing all colliding triangle pairs, we have a penetration depth energy vector $\text{PDe}\in\mathbb{R}^{|\mathcal{V}|}$. The overall computation is described by \prettyref{alg:pd}.
{
\begin{algorithm}
    \caption{Generating Penetration Energy Vector $\text{PDe}$ }
    \label{alg:pd}
    \begin{algorithmic}[1]
    {
    \small
    \State {Init $\text{PDe} = \Vec{0}\in\mathbb{R}^{|\mathcal{V}|}$}
    \State {Run FCL finding the set of all collided disjoint triangle paris as $\hat{T}$}
    \For {$(\mathbf{t}_p, \mathbf{t}_q) \in \hat{T}$ and the corresponding $\text{PD}_{p,q}$}
        \For {Vertex $i$ belongs to $\mathbf{t}_p$ and $\mathbf{t}_q$}
            \State {$\text{PDe}_i \mathrel{+}= {\text{PD}}_{p,q}^2/{6}$}
        \EndFor
    \EndFor
    }
\end{algorithmic}
\end{algorithm}
}

After computing the PDe, we use the following domain-decomposed data loss to train $\mathcal{S}_i$:
{
\small
\begin{align*}
\mathcal{L}_{PD}=\sum_{k=1}^{|\mathcal{Z}|_0}
\|\mathcal{S}_k-\sum_{i=1}^{|\mathcal{V}|}\mathcal{M}^{ki}\text{PDe}_i\|^2 + \|\mathcal{S}_{sum}-\text{PDe-sum}\|,
\end{align*}
}
where $\text{PDe-sum} = \sum_{i=1}^{|\mathcal{V}|}\text{PDe}_i$ is the ground truth total penetration energy and $\mathcal{S}_{sum} = \sum_{k=1}^{|\mathcal{Z}|_0}\mathcal{S}_k$ is the neural network prediction. Here, we use the same attention mask $\mathcal{M}^{ki}$ defined in \prettyref{sec:attention} to decompose the collision energy into body parts. Note that we do not have any loss terms related to $\mathcal{S}_0$. However, a neural network is known to suffer from over-fitting when learning exact distance functions~\cite{hoffer2015deep,burges2005learning}, including those corresponding to PD. Further, it is inherently difficult to train a perfect regression model for values like collision penetration depth with a long-tail distribution~\cite{wang2017learning}. We avoid over-fitting by using the marginal ranking loss. Given two meshes, $\mathcal{G}$ and $\hat{\mathcal{G}}$ (with approximated total penetration energy denoted as $\mathcal{S}_{sum}$ and $\widehat{\mathcal{S}_{sum}}$) randomly sampled from the dataset, if $\hat{\mathcal{G}}$ has a higher collision violation than $\mathcal{G}$ in terms of the total penetration energy, then we define:
{
\small
\begin{align*}
\mathcal{L}_{rank}=\E{max}(0,\alpha-(\widehat{\mathcal{S}_{sum}}-\mathcal{S}_{sum})),
\end{align*}
}
and vice versa. Here, $\alpha$ is used as a margin to enforce ranking strictness. We choose $\alpha $ as the mean energy difference of the given dataset. 

With the above training technique, we can predict $\mathcal{S}_1,\mathcal{S}_{|{Z}_0|}$ and use them as hard constraints by letting $\mathcal{S}_i=0$, resulting in $|{Z}_0|$ constraints. We can further reduce the online
computational cost by reducing the number of constraints to only one. To perform this computation, we train a single classifier $\text{MLP}_{classifier}(\mathcal{S}_1,\cdots,\mathcal{S}_{|{Z}_0|})$ to summarize the information and predict whether there are any collisions throughout the human body, i.e. $\text{MLP}_{classifier}$ is an indicator of whether $S_{sum} = 0$. To make sure that the $0.5$ sub-level set is the collision-free subset, we use the cross entropy loss:
{
\small
\begin{align*}
\mathcal{L}_{entropy}=&-\mathbb{I}(\text{PDe-sum}>0)\E{log}(\text{MLP}_{classifier})  \\
&-\mathbb{I}(\text{PDe-sum}=0)\E{log}(1-\text{MLP}_{classifier}).
\end{align*}
}

\subsection{\label{sec:OPT}Solving Constrained Optimization}
Our collision response solver takes a constrained optimization in the following form:
\begin{equation}
\begin{aligned}
\label{eq:OPT}
\argmin{\mathcal{Z}_{all}}&\|\mathcal{Z}_{all}-\mathcal{Z}_{all}^*\|^2 \\ 
\ST&\text{MLP}_{classifier}\THREE{\mathcal{S}_1}{\cdots}{\mathcal{S}_{|\mathcal{Z}_0|}}\leq 0.5.
\end{aligned}
\end{equation}
The idea is to provide a desired pose $\mathcal{Z}_{all}^*$ for the bilevel decoder, and \prettyref{eq:OPT} solves for a collision-free $\mathcal{Z}_{all}$ that is as close to $\mathcal{Z}_{all}^*$ as possible. We solve \prettyref{eq:OPT} using the augmented Lagrangian method implemented in LOQO \cite{vanderbei1999loqo}, with all the gradient information computed via back-propagation through the neural network. This augmented Lagrangian method can start from an infeasible domain, which means that LOQO allows the hard constraints to be temporarily violated between the iterations. As a result, LOQO uses gradient information to pull the solution back to the feasible sub-manifold.

\section{Evaluation}
We implement our method using PyTorch \cite{paszke2017automatic}. All the training and testing are performed on a single desktop machine with a $4$-core CPU, $32$GB memory, and an NVIDIA GTX 1080Ti GPU. The training is decomposed into two stages. During the first stage, we train the two-level human pose embedding architecture  using a set of $N$ meshes. This training would optimize only the $|\mathcal{Z}_0|+1$ autoencoders and the attention mechanics. After this first stage, we generate a much larger dataset of $M\gg N$ meshes by sampling the latent code $\mathcal{Z}_{all}$ uniformly in the range:
\begin{align*}
[1.2\E{min}({\mathcal{Z}_{all}}),1.2\E{max}({\mathcal{Z}_{all}})]^{|\mathcal{Z}_{all}|},
\end{align*}
where $\E{min}({\mathcal{Z}_{all}}_k)<0$, $\E{max}({\mathcal{Z}_{all}}_k)>0$, and $\E{min,max}$ are elementwise over all mesh samples.

We train our collision predictor and classifier on the augmented dataset while fixing the $|\mathcal{Z}_0|+1$ autoencoders and the attention mechanics. This stage uses the loss:
\begin{align*}
\mathcal{L}=w_{PD}\mathcal{L}_{PD}+w_{rank}\mathcal{L}_{rank}+w_{entropy}\mathcal{L}_{entropy},
\end{align*}
which is configured with $w_{PD}=5$, $w_{rank}=2$, $w_{entropy}=2$, and trained using a learning rate of $0.001$ and a batch size of $32$ over $30$ epochs. We evaluate our method on three datasets: the SCAPE dataset \cite{anguelov2005scape} with $N=71$ meshes, the MIT Swing dataset \cite{vlasic2008articulated} with $N=150$ meshes, and the MIT Jumping dataset \cite{vlasic2008articulated} with $N=150$ meshes. For each dataset, we use all the meshes to train the embedding space during the first stage, where we set $|\mathcal{Z}_0|=10$ for SCAPE and $|\mathcal{Z}_0|=12$ for Swing and Jumping. During the second stage, we use $0.7M$ samples of the augmented dataset for training and $0.3M$ samples for validation. We use two settings, one with $M=5\times10^4$ and the other with $M=2.5\times10^6$. 

\begin{table}[th]
\setlength{\tabcolsep}{3pt}
\begin{center}
\begin{tabular}{cccc}
\toprule
Baseline & MSE & RANK & CLASSIFY    \\
\midrule
$Ours$ & $6.72\times10^{-4}$ & $6.5\times10^{-3}$ & $82.8\%$  \\
$\mathcal{L}_{entropy}+\mathcal{L}_{PD}$ & $5.06\times10^{-4}$ & $9.4\times10^{-3}$ & $81.1\%$  \\
$\mathcal{L}_{entropy}+\mathcal{L}_{rank}$ & - & $4.7\times10^{-3}$ & $80.7\%$ \\
$\mathcal{L}_{entropy}$ & - & - & $80.4\%$   \\
$ND$ & $6.9\times10^{-4}$ & $6.7\times10^{-3}$ & $80.2\%$   \\
\bottomrule
\end{tabular}
\end{center}
\vspace{-5px}
\caption{\label{table:ablation}We compare our method (Ours) with 4 baselines: $\mathcal{L}_{entropy}+\mathcal{L}_{PD}$, $\mathcal{L}_{entropy}+\mathcal{L}_{rank}$, $\mathcal{L}_{entropy}$, and ND (no collision decomposition). For each method, we train on the smaller dataset with $M=5\times10^4$ meshes, and we compare their accuracy in terms of predicting penetration depth energies (MSE), ranking penetration depth energies (RANK), and classifying collision-free meshes (CLASSIFY). The result shows that our hybrid loss improves the overall accuracy of collision predictions. Especially, the improvement over $ND$ demonstrates the effectiveness of decomposing a collision into body parts.}
\vspace{-10px}
\end{table}

\TE{Accuracy of Collision Prediction:} We consider several baselines that are essentially simplified variants of our pipeline in \prettyref{fig:pipeline}. We notice that the constrained optimization \prettyref{eq:OPT} only needs the output of $\text{MLP}_{classifier}$ to be correct, which is the goal $\mathcal{L}_{entropy}$. Therefore, we consider retaining only $\mathcal{L}_{entropy}$ while removing $\mathcal{L}_{rank}$ and $\mathcal{L}_{PD}$, leading to three baselines: $\mathcal{L}_{entropy}+\mathcal{L}_{PD}$, $\mathcal{L}_{entropy}+\mathcal{L}_{rank}$, and $\mathcal{L}_{entropy}$, where we use the same weights for the retained terms. In order to demonstrate the power of collision decomposition, we also compare our LCollision with a simplified network architecture that does not decompose the collision into body parts. For this baseline, we simply use $S_0$ to predict total penetration energy $\mathcal{S}_{sum}$ and classify collision status, and we modify $\mathcal{L}_{PD}$ to only have $\|\mathcal{S}_{sum}-\text{PDe-sum}\|$. The other two losses $\mathcal{L}_{PD}$ and $\mathcal{L}_{entropy}$ remain the same. This baseline is denoted as $ND$ (no decomposition).

In \prettyref{table:ablation}, we compare the accuracy of baselines in terms of predicting penetration depth energies, ranking penetration depth energies, and classifying collision-free meshes. To ensure that our predicted penetration depth energies are accurate, we use the mean squared error (MSE) of total penetration depth energy averaged over the $0.3M$ test meshes. To ensure the accuracy of the ranking penetration depth energies, we randomly formulate a pair for each sample in the $0.3M$ test meshes, and we record the average ranking margin (RANK). To classify collision-free meshes, we use the rate of success (CLASSIFY) over the $0.3M$ test meshes. 

From this ablation study, we compare $ND$ and our method to find that penetration decomposition can improve the accuracy of collision predictions, which also suggests that using the two parts of the $\mathcal{L}_{PD}$ could better inform the network of collision locality. Using penetration depth energy in the system not only provides gradient information for optimization but can also boost performance through $\mathcal{L}_{PD}$. 
$\mathcal{L}_{rank}$ does help improve performance, but its effect is relatively minor compared to $\mathcal{L}_{PD}$.
\begin{table}[ht]
    \setlength{\tabcolsep}{5pt}
    \begin{center}\scalebox{0.9}{
    \begin{tabular}{ccccc}
    \toprule
    $M$ & Dataset & MSE & RANK & CLASSIFY    \\
    \midrule
    \multirow{3}{*}{$5\times10^4$} & SCAPE & $6.72\times10^{-4}$ & $6.5\times10^{-3}$ & $82.8\%$  \\
        & Swing & $7.27\times10^{-4}$ & $3.38\times10^{-3}$ & $91.2\%$  \\
        & Jumping & $6.74\times10^{-4}$ & $5.29\times10^{-3}$ & $91.6\%$  \\
    \multirow{3}{*}{$2.5\times10^6$} & SCAPE & $7.80\times10^{-4}$ & $2.60\times10^{-3}$ & $94.1\%$  \\
        & Swing & $2.57\times10^{-4}$ & $2.34\times10^{-3}$ & $96.2\%$  \\
        & Jumping & $6.34\times10^{-4}$ & $5.43\times10^{-3}$ & $95.4\%$  \\
    \bottomrule
    \end{tabular}}
    \end{center}
    \vspace{-7px}
    \caption{\label{table:datasize} We study the robustness of our method in terms of dataset sizes. Increasing the dataset size $M$ can significantly boost the collision detection accuracy (CLASSIFY). This result implies that learning to predict collisions is challenging, and a larger training dataset can help improve the overall results. }
    \vspace{-10px}
\end{table}

\begin{table}[ht]
\setlength{\tabcolsep}{5pt}
\begin{center}
\begin{tabular}{cccccc}
\toprule
\multirow{2}{*}{Dataset} & \multirow{2}{*}{Time (Pan et al.)} & \multicolumn{2}{c}{Time (ours)} & \multicolumn{2}{c}{Speedup}    \\
&&CPU & GPU &CPU & GPU\\
\midrule
    SCAPE & 1min 23s & $3.99$s&$1.02$s & 21x &81x  \\
    Swing & 5min 12s & $3.78$s & $0.91$s & 82x & 342x  \\
    Jumping & 5min 19s & $4.23$s & $1.13$s & 75x& 282x  \\
\bottomrule
\end{tabular}
\end{center}
\vspace{-5px}
\caption{\label{table:time} We compare LCollision with \cite{pan2012fcl} in terms of the computational cost for collision detection. \cite{pan2012fcl} only supports the CPU version, while we tested both the CPU and the GPU versions of our method. All datasets have $1.5\times 10^4$ samples ($0.3M$ validation samples for $M=5\times 10^4$). Meshes in the Swing and Jumping datasets have more vertices ($9971$ and $10002$) than SCAPE ($2261$), and the complexity of \cite{pan2012fcl} depends on the number of points; thus, exact collision checking~\cite{pan2012fcl} takes more time. However, they all share the same level of latent space size with SCAPE, and the running times of our method are almost identical.}
\vspace{-10px}
\end{table}
Our second study inspects the robustness of our network architecture in terms of the size of the dataset. As shown in \prettyref{table:datasize}, we tested our method trained using two different $M$. Increasing $M$ from $5\times10^4$ to $2.5\times10^6$ can significantly boost the collision detection accuracy (CLASSIFY). This result implies that learning to predict collisions is challenging, and a larger training dataset can help improve the overall results. 

\TE{Speedup Compared with Exact Collision Checking:} The goal of our method is to speed up the collision detection process over prior, exact methods that are applied to mesh-based representations. We compare the running time with \cite{pan2012fcl} on the test set of $5\times 10^4$ samples ($1.5\times 10^4$ samples) for the SCAPE, Swing, and Jumping datasets. The implementation of \cite{pan2012fcl} only supports CPU, while LCollision runs on both CPU and GPU. To achieve the best performance for \cite{pan2012fcl}, we run their method using 15 threads in parallel and stop when one collision occurs or the mesh is reported to be collision-free. For our method, we feed the network with $500$ models at the same time. We optimize the hyper-parameters to obtain optimal performance. We show the results in \prettyref{table:time} and observe two orders of magnitude speedup.

\begin{figure}
\centering
\includegraphics[scale=0.4]{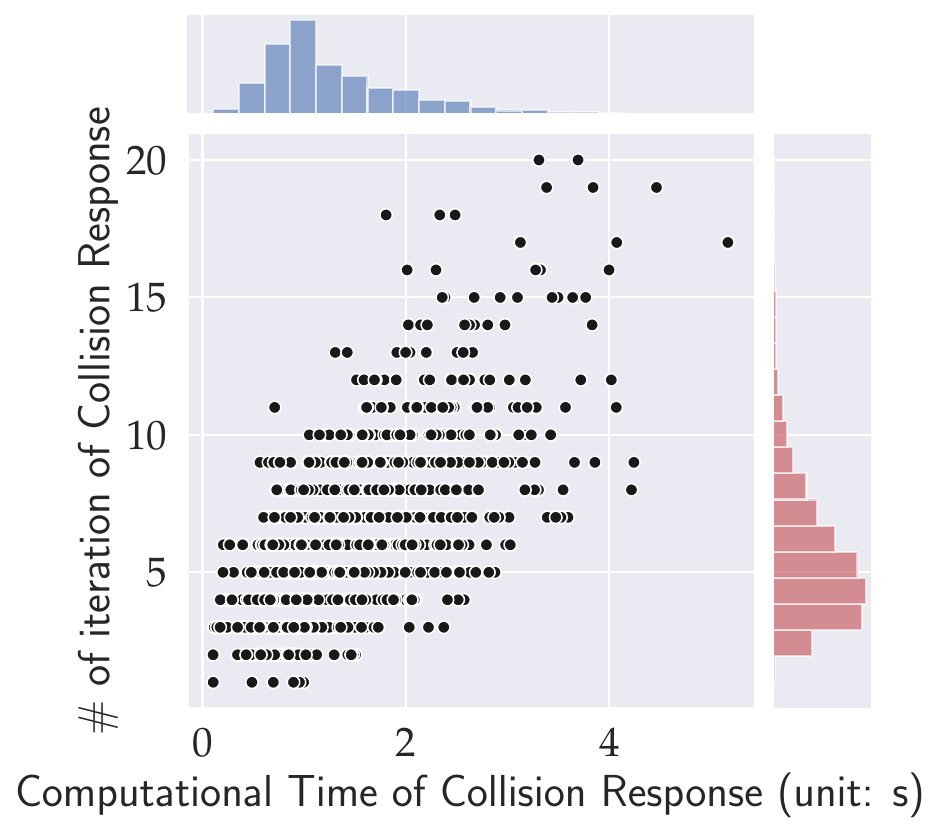}
\vspace{-10px}
\caption{\label{fig:convergence} The joint histogram of the number of iterations (Y-axis) and the computational time (X-axis) used for solving the constrained optimization (\prettyref{eq:OPT}) for the Swing dataset. The average number of iterations is 5.44 and the average computation time is 1.29s.}
\end{figure}

\begin{figure}
\centering
\includegraphics[scale=0.4]{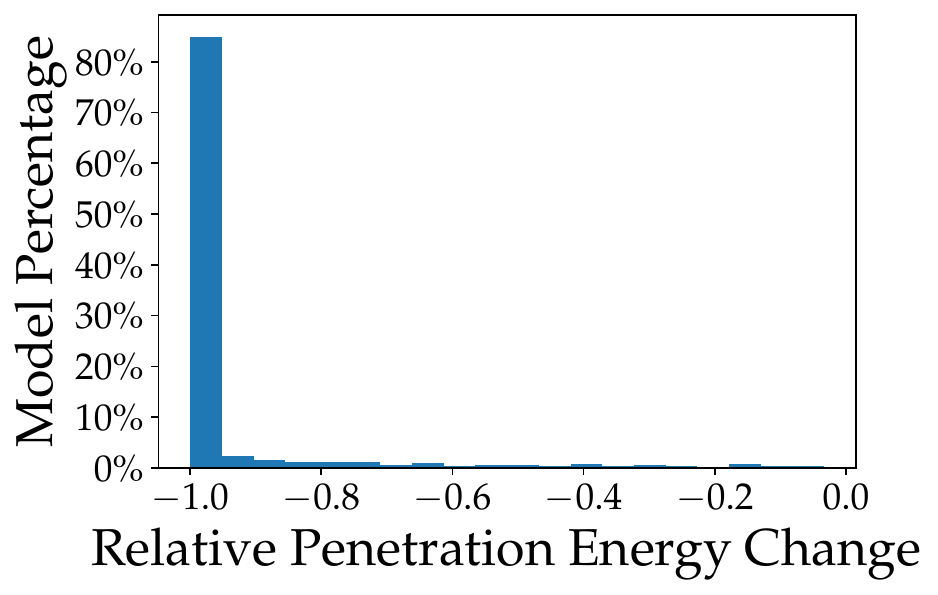}
\caption{\label{fig:response} The histogram of relative penetration energy change for successful examples in the Swing dataset. Our method achieves a success rate of $85.1\%$, and we observed an average relative decrease of $94.3\%$ in penetration energy.}
\end{figure}

\TE{The Collision Response Solver:} In \prettyref{fig:all}, we show 20 results with successful collision responses for the SCAPE dataset (more results on the other datasets given in the appendix). To profile the collision response solver quantitatively, we sample a set of 3000 random human poses by randomizing $\mathcal{Z}_{all}$ for both the SCAPE and Swing datasets. Some of the models have self-collisions and are classified correctly by our  learning-based collision detection algorithm. For each of these meshes, we solve \prettyref{eq:OPT} and we consider a solution successful if the augmented Lagrangian algorithm returns a feasible solution. On the SCAPE dataset, our method achieves a success rate of $85.6\%$, and we observe a relative decrease of $80.9\%$ in penetration energy. On the Swing dataset, our method achieves a success rate of $85.1\%$, and we observe a relative decrease of $94.3\%$. In \prettyref{fig:convergence}, we plot the number of iterations and computational time used by the constrained optimizer until convergence for the Swing dataset. The average iteration is 5.44 and the average time is 1.29s. For the SCAPE dataset, the average iteration is 2.09 and the average time is 0.25s. In \prettyref{fig:response}, we highlight the distribution of relative penetration energy change for successful collision response models.

\section{Conclusion, Limitations, and Future Work}
We present LCollision, a method for learning the collision-free human pose sub-manifold. We use a mesh embedding autoencoder to learn a full human pose manifold and augment it with additional components to classify the collision-free meshes. Our method decomposes the mesh into several sub-domains and learns the decision boundary of the collision-free sub-manifold by reusing the decomposed sub-domains. Specifically, we learn to predict the penetration depths aggregated to each sub-domain and then use a binary classifier to predict whether a given mesh has any collisions. When evaluated on the SCAPE dataset, our method achieves a success rate of $94.1\%$ in predicting collisions and a success rate of $85.6\%$ in collision responses. 

Our method has some limitations. Being a learning-based method, our collision predictor cannot achieve a $100\%$ success rate, in contrast to exact collision detection algorithms. This could pose a problem when our method is used to generate computer animations, where a few missed collisions can have a considerable impact on the overall simulation accuracy. Moreover, our learning method can only be applied to models with fixed topology and requires additional data collection and training for different mesh topologies. In the future, we would like to consider active learning to collect more data and improve the accuracy of the collision predictor in a self-supervised manner, and thereby reduce the need for large training datasets. A similar approach is used in~\cite{pan2013efficient,he2015efficient} for rigid objects. A second issue is the use of a continuous constraint optimizer \cite{vanderbei1999loqo} for collision responses. These solvers require twice-differentiable hard constraints, which is not the case in our application because we use non-differentiable ReLU activation units. It is worth exploring new constraint optimization solvers that could work with non-smooth constraints specified by a neural network. There are many issues in terms of incorporating hard constraints into a neural network.
If only soft penalties are needed, we can reformulate the hard constraint in \prettyref{eq:OPT} as a soft penalty term and solve the unconstrained problem via a Newton-Type method, allowing users to adjust the penetration allowed in the final. We can extend our work by considering other types of hard constraints such as dynamics and accurate collision response models. Finally, since our method uses a hybrid loss, it may compromise the performance in some metrics, e.g., the regression loss of the penetration depth. Moreover, techniques based on parameter estimation can be used to improve the performance of such learning methods~\cite{wolinski2014parameter}.

\section*{Acknowledgements}
This work was supported in part by ARO under Grants W911NF1810313 and W911NF1910315, and in
part by Intel.
\bibliography{references.bib}
\end{document}